# High-Resolution Spectroscopy in the 11.6–15 µm Range by a Quasi-cw Difference-Frequency-Generation Laser Source

Ali Elkhazraji,[1,3] Mohammad Khaled Shakfa,[1,3] Marco Lamperti,[2] Khaiyom Hakimov,[1] Khalil Djebbi,[1] Riccardo Gotti,[2] Davide Gatti,[2] Marco Marangoni,[2,4] and Aamir Farooq[1,*]

[1]King Abdullah University of Science and Technology (KAUST), Clean Combustion Research Center, Physical Sciences and Engineering Division, Thuwal 23955-6900, Saudi Arabia
[2]Politecnico di Milano, Department of Physics and IFN-CNR, Via G. Previati 1/C, 23900 Lecco, Italy
[3]Ali Elkhazraji and Mohammad Khaled Shakfa contributed equally to this work
[4]marco.marangoni@polimi.it
*aamir.farooq@kaust.edu.sa

**Abstract:** We report an approach for high-resolution spectroscopy using a widely tunable laser emitting in the molecular fingerprint region. The laser is based on difference-frequency generation (DFG) in a nonlinear orientation-patterned GaAs crystal. The signal laser, a $CO_2$ gas laser, is operated in a kHz-pulsed mode while the pump laser, an external-cavity quantum cascade laser, is finely mode-hop-free tuned. The idler radiation covers a spectral range of ~11.6 – 15 µm with a laser linewidth of ~ 2.3 MHz. We showcase the versatility and the potential for molecular fingerprinting of the developed DFG laser source by resolving the absorption features of a mixture of several species in the long-wavelength mid-infrared. Furthermore, exploiting the wide tunability and resolution of the spectrometer, we resolve the broadband absorption spectrum of ethylene ($C_2H_4$) over ∼13 – 14.2 µm and quantify the self-broadening coefficients of some selected spectral lines.

## 1. Introduction

Mid-infrared (mid-IR) spectroscopy is a powerful tool for molecular fingerprinting: it offers a unique capability to identify and quantitatively determine chemical species with high sensitivity and resolution. Several spectroscopic techniques have been utilized for molecular fingerprinting. Fourier transform infrared (FTIR) spectrometry based on broadband thermal light sources has been a golden standard for years. It offers an ultra-broadband spectral coverage, but it exhibits a spectral resolution limited by the scanning range of the moving arm of the interferometer. Only with very long arms and bulky setups can the resolution be brought to the 0.001 $cm^{-1}$ level needed to resolve gas absorption features of small molecules with minimal instrumental broadening [1]. To overcome this issue, optical frequency combs can be used as a light source for FTIR spectroscopy [2–4]. Optical frequency combs can be employed for direct absorption spectroscopy to perform molecular fingerprinting with extreme frequency precision [5,6] over broad bands. The spectra are composed of up to several hundred thousand points evenly separated by the comb repetition rate, typically 100 to 250 MHz for combs derived from mode-locked oscillators. The spectral resolution of each point depends on the linewidth of the individual comb modes and can be shrunk to the Hz level for highly coherent frequency combs [5,6]. On the other hand, to fill the spectral gaps between adjacent points, the frequency comb must be either shifted or tuned like an accordion, which implies changing its offset or its repetition frequency, respectively. Interestingly, FTIR spectroscopy with frequency combs at high resolution can be performed by replacing the bulky mechanical delay arm of the Michelson interferometer with a second frequency comb operated at a slightly different frequency spacing, according to the so-called dual-comb spectroscopy approach [7,8]. This was shown to be very powerful for multispecies trace gas sensing [9] in the atmosphere and for the study of transient processes such as chemical reactions [10] and combustion processes [11]. The synthesis of frequency combs in the mid-IR remains a nontrivial task, however, as witnessed by the absence of a clearly winning solution after almost 15 years of developments and achievements [12]. The general approach implies nonlinear frequency conversion of near-infrared lasers, but depending on the addressed part of the mid-IR spectrum and on the targeted bandwidth and tunability, different processes with different nonlinear crystals may be implemented, mostly based on optical-parametric-oscillation (OPO) [13–15] and difference-frequency-generation (DFG) [16–18], also in the intra-pulse configuration [19,20].

The advancements in frequency comb spectroscopy do not remove the need for continuously tunable lasers to spectrally resolve, with a large number of points, narrow lines in the mid-IR region, particularly at wavelengths beyond 10 μm where the Doppler width and the typical line densities are consistently below the spacing between adjacent comb modes. In this area, semiconductor diode lasers do not offer wavelengths > 5 μm [21] while fiber lasers are limited to about 3.5 μm [22]. In contrast, quantum cascade lasers (QCLs) have been extensively developed in recent years for longer mid-IR wavelengths and may offer mode-hop-free tuning, narrow linewidth, and high output power [23–25]. However, cw-QCLs suffer in general from a very limited wavelength tunability and, at least so far, from a scarce availability of commercial sources beyond 13 μm.

Here, we present an approach for high-resolution molecular spectroscopy in the long-wavelength part of the mid-IR region that exploits a difference-frequency generation (DFG) process between an external-cavity quantum cascade laser (EC-QCL, pump) and a $CO_2$ gas laser (signal) to produce quasi-cw laser radiation tunable without gaps from ~11.6 to 15 μm. This laser system was previously applied in a fixed-wavelength modality for environmental monitoring and combustion applications [26–29]. It was also used for optical metrology of bending modes, in combination with a frequency comb and with a distributed-feedback-QCL in substitution of the EC-QCL, though at the price of a narrower fine tunability [30]. In this work we fully exploit, instead, the very large mode-hop-free tuning range of the EC-QCL, over more than 60 $cm^{-1}$, to acquire gap-free ultra-broad absorption spectra from 666.67 to 863.56 $cm^{-1}$ (11.58 – 15 μm) with few-MHz resolution.

## 2. DFG laser: motivation, design, and characterization

The spectral region spanning 11.6 – 15.0 μm (∼670 – 860 $cm^{-1}$) is a particularly interesting region for molecular spectroscopy as it entails dense and strong fundamental vibrational bands of a multitude of molecules. Though it covers only ∼3.2% of the spectral range covered by the PNNL database (an empirical spectral database of absorption cross-sections of the most common molecules for remote sensing and environmental applications over ∼600 – 6500 $cm^{-1}$), about 19% of the species listed in the database (92 out of ∼480) have their strongest IR bands in this spectral range (11.6 – 15 μm) [31]. BTEX (benzene, toluene, ethylbenzene, and *m,o,p*–xylene isomers) species and halocarbons are two interesting families of such species, showing their strongest IR absorption bands in this spectral range. Figure 1(a) shows the absorbance spectra of BTEX retrieved from the PNNL database [31] across the entire mid-IR region. Due their similar chemical structure, their spectra are strongly overlapped both in the CH stretching region at 3 – 3.5 μm (lower figure inset) and in the bending mode region at 12 – 15 μm (upper inset), but in the latter region the absorption features of the different compounds are by far more distinguishable. This makes the long-wavelength mid-IR the range of election for selective and quantitative sensing of BTEX compounds for environmental monitoring, as BTEX are among the most concerning pollutants [32]. On another front, halocarbons are ozone-depleting substances (commonly known for refrigeration applications) with a global warming potential (GWP100) three-to-four orders of magnitude greater than that of $CO_2$ [33]. The strongest IR bands of many halocarbons (C–halogen stretching modes) exist between 11 and 14 μm, overlapping the peak of terrestrial radiation and the transparent atmospheric window (∼8 – 14 μm), where atmospheric absorption reaches a minimum due to the low-to-no absorption of $CO_2$ and $H_2O$ vapor [34]. This makes halocarbons potent greenhouse gases blocking significant portions of the atmospheric window. Figure 1(b) shows the absorbance spectra of selected halocarbons in the IR atmospheric window, as delimited by the strong absorption bands of $CO_2$ and $H_2O$ vapor. They are congested yet susceptible for being tracked and discriminated in this region. This is of motivation for the realization of a broadly tunable laser source in this spectral region (11 – 15 μm), not only for quantitative detection of halocarbons but also to determine accurate spectroscopic parameters and refine global warming and radiative efficiency calculations [33,35].

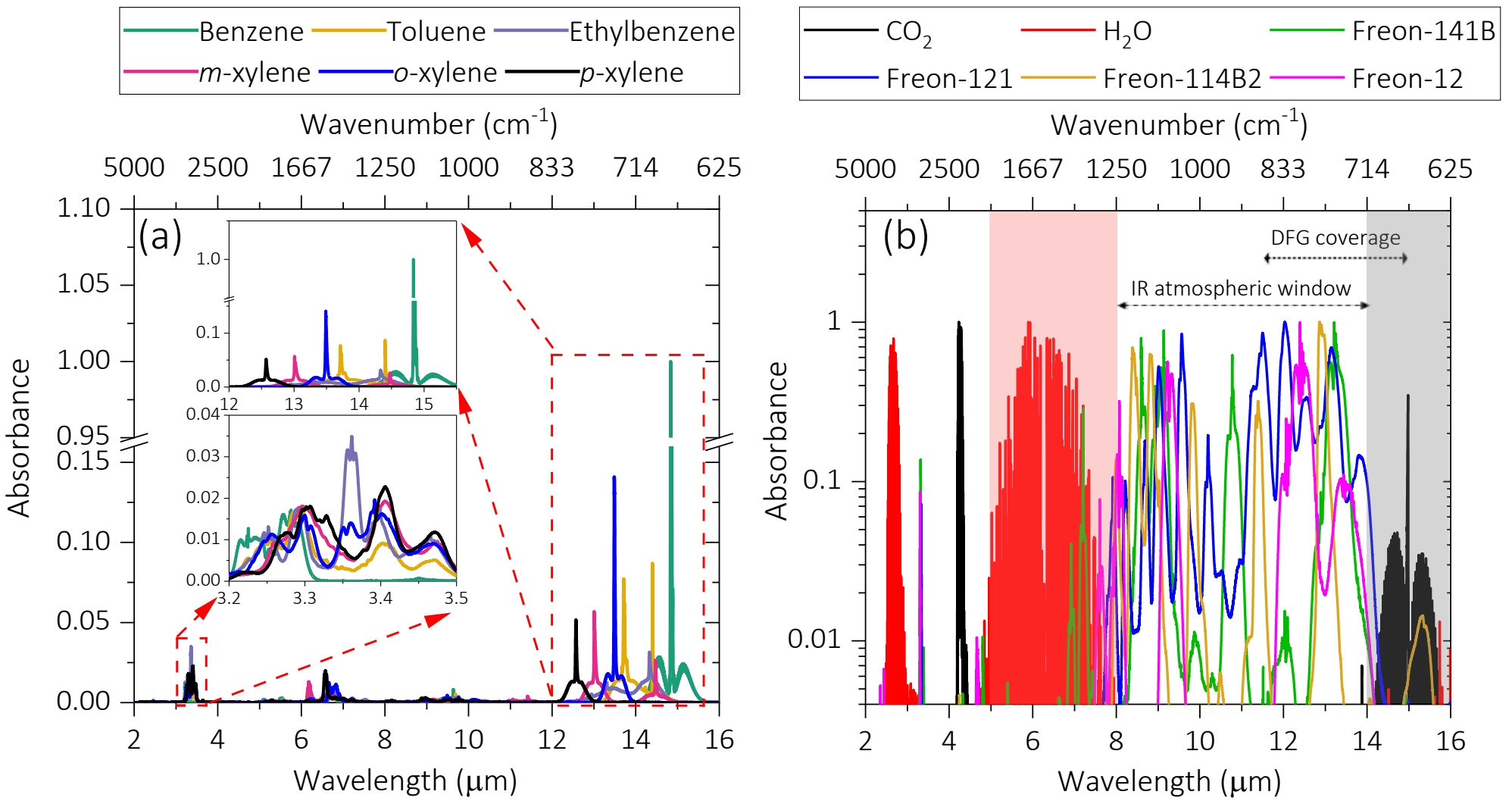

Fig. 1. Normalized absorbance spectra in the IR region at 1 atm, 298 K retrieved from the PNNL database [31] of (a) BTEX, with insets showing close-up views of the CH stretching vibrational modes (lower inset) and CH bending modes (upper inset) and (b) selected halocarbons; the red and black highlighted regions define the edges of the IR atmospheric window bounded by strong $H_2O$ and $CO_2$ absorption bands, respectively.

A schematic of the developed mid-IR DFG laser source is shown in Fig. 2(a). An EC-QCL (*Daylight Solutions*, *41058-MHF model*) with single-mode emission between 5.45 and 5.71 μm (1749.97 – 1834.93 cm$^{-1}$) is used as a pump laser. The signal laser is a pulsed $CO_2$ gas laser (*Access Lasers*, *L20GD model*) that offers several lasing lines between 9.23 and 10.86 μm (920.81 – 1083.42 cm$^{-1}$). Pump and signal lasers exhibit linewidths of 1.7 MHz [36] and 1.5 MHz [30], respectively, as measured through the beating with an optical frequency comb. Both lasers emit vertically polarized radiation. We installed an optical isolator (OI) in front of the EC-QCL to suppress back reflections that may affect or damage the laser, mostly those from the facets of the OP-GaAs crystal used for DFG. The OI causes a rotation by 45° of the pump laser polarization that is compensated by a half wave-plate (WP). This allows selecting an optimized polarization state for the pump field in the nonlinear crystal and maximizing the conversion efficiency.

We used two sets of concave mirrors to attain spot sizes that optimize the spatial overlap between pump and signal laser beams in the nonlinear crystal. Pump and signal beams are superimposed by a beam combiner and focused onto the OP-GaAs crystal by a parabolic mirror ($PM_1$). A second parabolic mirror ($PM_2$) collects and collimates the idler beam after the OP-GaAs crystal. We used a reflective long-pass filter with a cut-on edge near 11.2 μm to filter out (reflect) residual pump and signal lasers before the gas cell.

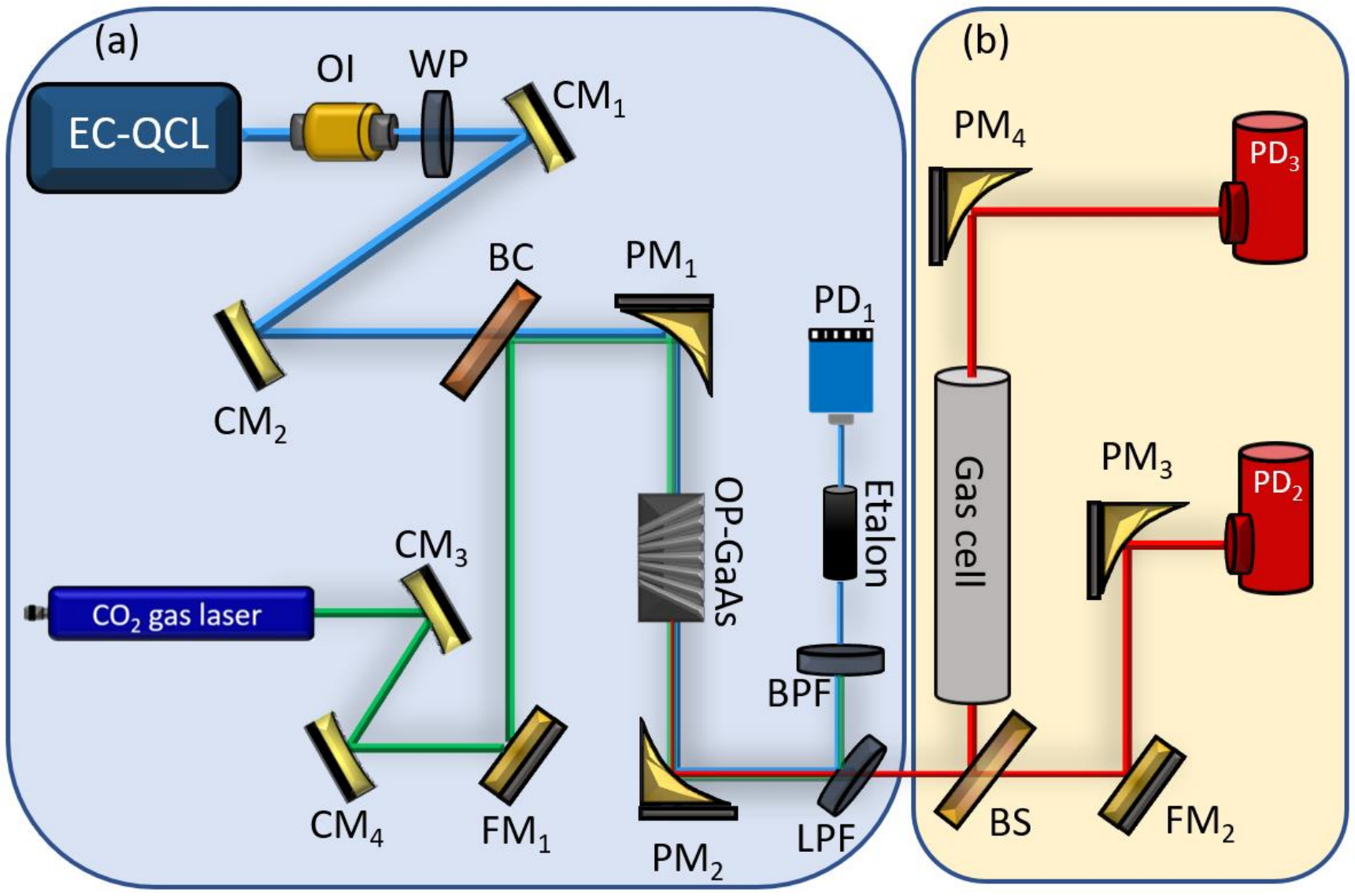

Fig. 2. Schematic of the widely tunable mid-infrared spectrometer. (a) Difference frequency generation laser under nitrogen purging. (b) Gas testing setup with a 23-cm-long optical cell and two $LN_2$-cooled HgCdTe detectors. EC-QCL: external-cavity quantum cascade laser; OI: optical isolator; WP: wave plate; CM: concave mirror; FM: flat mirror; BC: beam combiner; PM: parabolic mirror; OP-GaAs: orientation-patterned GaAs crystal; LPF: long-pass filter; BPS: bandpass filter; PD: photodetector; BS: beam splitter.

Figure 3(a) shows the power-vs-wavelength behavior of the pump laser (EC-QCL) measured with a photodetector at a distance of about 20 cm from the laser aperture in ambient air. On a zoomed-in wavelength axis (see figure inset) the emitted power is affected by oscillations of two different scales, a faster one with 0.5% amplitude and a slower one with 2.5 % amplitude, which most likely arise from etalons within the EC-QCL itself. The sharp features across the spectrum are instead due to absorption lines of atmospheric water vapor. To minimize the effect of humidity on the output power and on spectroscopy measurements, we placed the laser setup in a sealed enclosure purged with nitrogen which reduced the relative humidity in the laser box to ~ 0.7% from the ambient value of ~ 60%. Figure 3(b) shows representative DFG average power curves as a function of the idler wavelength at a few selected wavelengths of the signal laser ($CO_2$ gas laser), as operated at a repetition rate of 2.5 kHz with a duty cycle of 30%. We obtained these curves by setting the signal laser to a specific wavelength while stepwise changing the pump wavelength over the entire tuning range. The position of the OP-GaAs crystal was adjusted at each pump wavelength to optimize the DFG (idler) output power. The idler wavelength can be tuned between 11.58 μm (863.56 $cm^{-1}$) and 15.00 μm (666.67 $cm^{-1}$), with the highest output power of about 31 μW (100 μW peak power) near 12.63 μm. The idler laser linewidth of ~2.3 MHz was estimated from the convolution of the pump and signal laser linewidths.

The OP-GaAs crystal used in the DFG process is 35 mm long and realized with a fan-out structure with poling periods from 183 to 203 μm. This ensures phase matching over the entire tuning range of the pump laser. The OP-GaAs crystal was designed to have a width of 20 mm, which is sufficiently large to avoid any efficiency drop due to non-uniform phase-matching conditions between the pump and signal laser beams. The conversion efficiency changes while tuning the pump wavelength according to the typical $\text{sinc}^2$ behavior [37] with a phase-matching bandwidth (full width at half maximum) of about 20 nm corresponding to ~6 $cm^{-1}$, as shown in Fig. 3(c). This implies that there is no need to adjust the crystal position while tuning the pump wavelength, at a specific signal wavelength, when scanning within a wavenumber range of 6 $cm^{-1}$ (or broader, depending on the signal quality outside the phase-matching bandwidth).

The fluctuations in the phase-matching curve of Fig. 3(c) arise from the intrinsic power oscillations of the pump laser and from residual water-induced absorption of the pump radiation inside the DFG laser enclosure, which persisted even upon nitrogen purging.

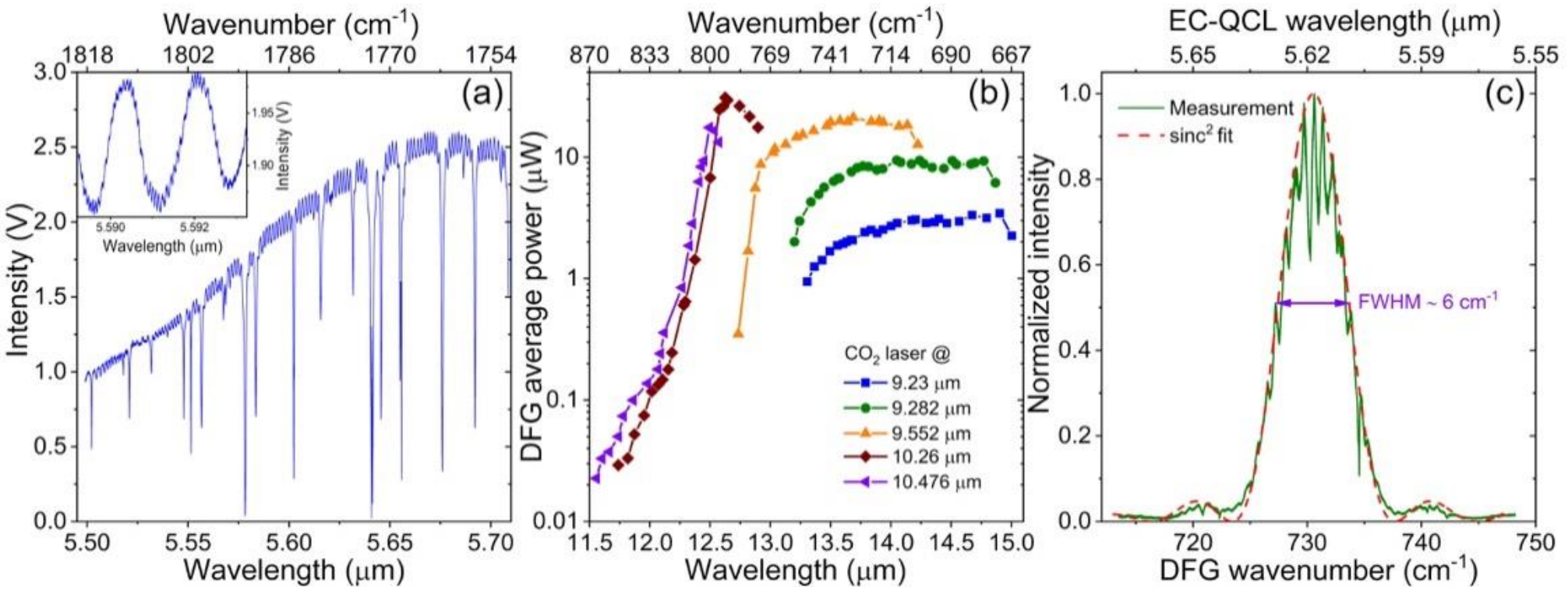

Fig. 3. (a) EC-QCL power tuning curve measured at a distance of ~20 cm from the laser aperture in ambient air. The sharp features across the spectrum are due to atmospheric water absorption lines. The inset displays two scales of oscillations in the EC-QCL spectrum. (b) Average power curves of the DFG laser measured at various $CO_2$ gas laser wavelengths when the EC-QCL wavelength is stepwise tuned. The DFG wavelength varies between 11.58 μm (863.56 cm$^{-1}$) and 15.00 μm (666.67 cm$^{-1}$). (c) Phase-matching curve (normalized) of the DFG mixing process measured at a fixed $CO_2$ gas laser wavelength of 9.52 μm. The dashed line shows a sinc$^2$ fitting.

Figure 2(b) shows the measurement part of the setup equipped with two liquid-nitrogen-cooled HgCdTe detectors (*Infrared Associates Inc.*). The idler laser beam is split into two beams: one is directly focused onto the first detector and used as a reference, while the other is aligned to a 23-cm-long gas cell equipped with coated-ZnSe windows and then focused on to the second detector. We used an oscilloscope (*PicoScope 5000*, *Pico Technology*) to record detector signals, which were post-processed and analyzed using an in-house MATLAB code.

The idler wavelength is tuned by scanning the EC-QCL wavelength while fixing that of the $CO_2$ gas laser. The wavelength change of the EC-QCL is obtained through mechanical motorized rotation of its diffraction grating. During frequency scans, a wavelength reference TTL pulse generated by the EC-QCL every 0.2 nm is recorded alongside the detector signals. Due to the nonuniformity in the tuning speed within each 0.2 nm spectral segment (i.e., in between two consecutive EC-QCL TTL pulses), any linear interpolation of the spectral points between TTL pulses resulted in a distorted (stretched or compressed) frequency axis. To solve the issue, a silicon Fabry–Pérot etalon (*LightMachinery*) with a free spectral range (FSR) of 0.0175 cm$^{-1}$ was added to the setup for a more accurate calibration of the pump wavelength. As shown in Fig. 2(a), the pump beam is aligned to the Fabry–Pérot etalon after the removal of signal and idler beams by a sequence of long-pass and band-pass filters (BPF). The Fabry–Pérot throughput is measured by a photodetector ($PD_1$) in a synchronous way with the idler intensities measured by $PD_2$ and $PD_3$ at the cell input and output, respectively. We used the line center values of ethylene absorption lines in the HITRAN2020 database [38] to obtain an absolute calibration of the wavelength axis. The wavelength calibration could also be confirmed by monitoring the water absorption lines in the EC-QCL tuning curve. Figure 4(a) shows an example of raw signals recorded during the measurement of an ethylene ($C_2H_4$) absorption spectrum at 295 K and 5 Torr. Here, the $CO_2$ gas laser wavelength was set to 9.552 μm with a pulse repetition rate of 10 kHz while the EC-QCL wavelength was scanned over 5.5916 – 5.5982 μm. Figures 4(b) and (c) report, for a selected spectral segment, the laser intensity oscillations and the corresponding fluctuation of the calculated absorbance, respectively.

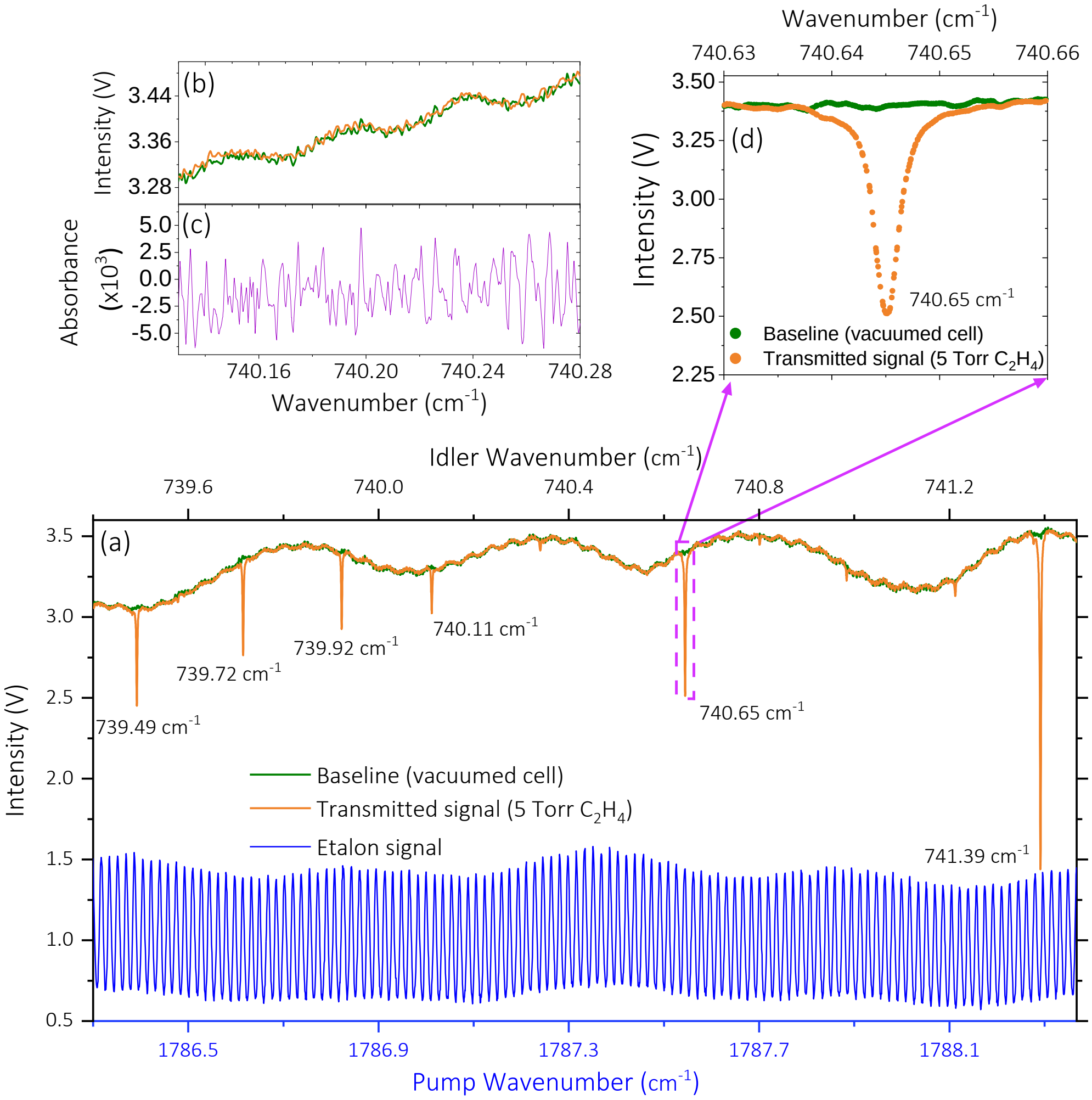

Fig. 4. (a) Measured spectra at the cell output when the cell is under vacuum (green baseline), when the cell is filled with $C_2H_4$ at a pressure of 5 Torr and temperature of 295 K (8 interlaced scans, orange line), as compared to the etalon signal registered at the pump wavelength; the shown label next to each line denotes its linecenter. The bottom axis corresponds to the scanned wavelength of the pump laser (EC-QCL, for the etalon signal) while the top axis gives the corresponding idler wavelength. (b) Fluctuations of the spectral baseline in a given spectral portion. (c) Corresponding fluctuation in terms of absorbance, for the same spectral portion. (d) Close-up view of a single $C_2H_4$ line centered at 740.65 cm$^{-1}$ showing the baseline and transmitted signals.

For the spectrum illustrated in Fig. 4(a), the scan speed of the EC-QCL was set to the minimum value of 25 nm/s to increase the number of spectral points, which corresponds to the number of signal laser pulses in the measurement time. In other words, the spectral spacing between successive data points can be reduced by decreasing the scan speed of the pump laser and/or by increasing the pulse repetition rate of the signal laser. By repeating the measurement one can further reduce the spectral spacing and densely resolve even narrow absorption features: this is shown in Fig. 4(d) for the spectral line centered at 740.65 cm$^{-1}$, whose spectrum is composed of eight interlaced scans with the $CO_2$ laser pulse repetition rate set to 10 kHz. The effectiveness of repeating the scan measurements to reduce the spectral spacing originates from the scarce repeatability of EC-QCL scans and from the absence of synchronization between signal laser pulse sequence and starting time of consecutive EC-QCL scans. Hence, the DFG process between the train of $CO_2$ laser pulses and the cw EC-QCL takes place arbitrarily in time for repeated measurements.

## 3. High-resolution mid-IR molecular fingerprinting

To showcase the potential of the developed DFG laser system for molecular spectroscopy, we carried out a high-resolution measurement of the absorption spectrum of a non-diluted multispecies mixture composed of ethylene ($C_2H_4$), carbon dioxide ($CO_2$), ammonia ($NH_3$) and acetylene ($C_2H_2$) over 740.5 – 748.5 cm$^{-1}$. The signal laser ($CO_2$ gas laser) wavelength was set to 9.552 μm at a repetition rate of 10 kHz, while the pump wavelength was scanned over 5.5698 – 5.5947 μm. Figure 5(a) shows the measured spectrum of the gas mixture at 50 Torr along with the simulated spectra using HITRAN2020 database [38]. The integrated absorbances of the observed lines were used to infer the mole fractions of the species in the mixture, using the integrated form of the Beer-Lambert Law:

$$x_i = \frac{A_i}{S_i P L} \tag{1}$$

where $x_i$ is the mole fraction of the $i^{\text{th}}$ absorbing species, $A_i$ is the integrated absorbance of the targeted spectral transition, $S_i$ is the line strength of the transition (retrieved from the HITRAN2020 database [38]), $P$ is the pressure, and $L$ is the pathlength. The obtained least-squared mole fraction values were then used to simulate the absorbance spectra of the individual species in the mixture (shown in the bottom panels of Fig. 5(a) and (b)), and a composite spectrum was generated by adding up the individual simulated spectra (black dashed lines in the upper panels of Fig. 5(a) and (b)). Experimental and simulated spectra are in good agreement, though the accuracy of the mole fraction determination was impaired by the strong adsorption of $NH_3$ to the walls of the stainless-steel static cell.

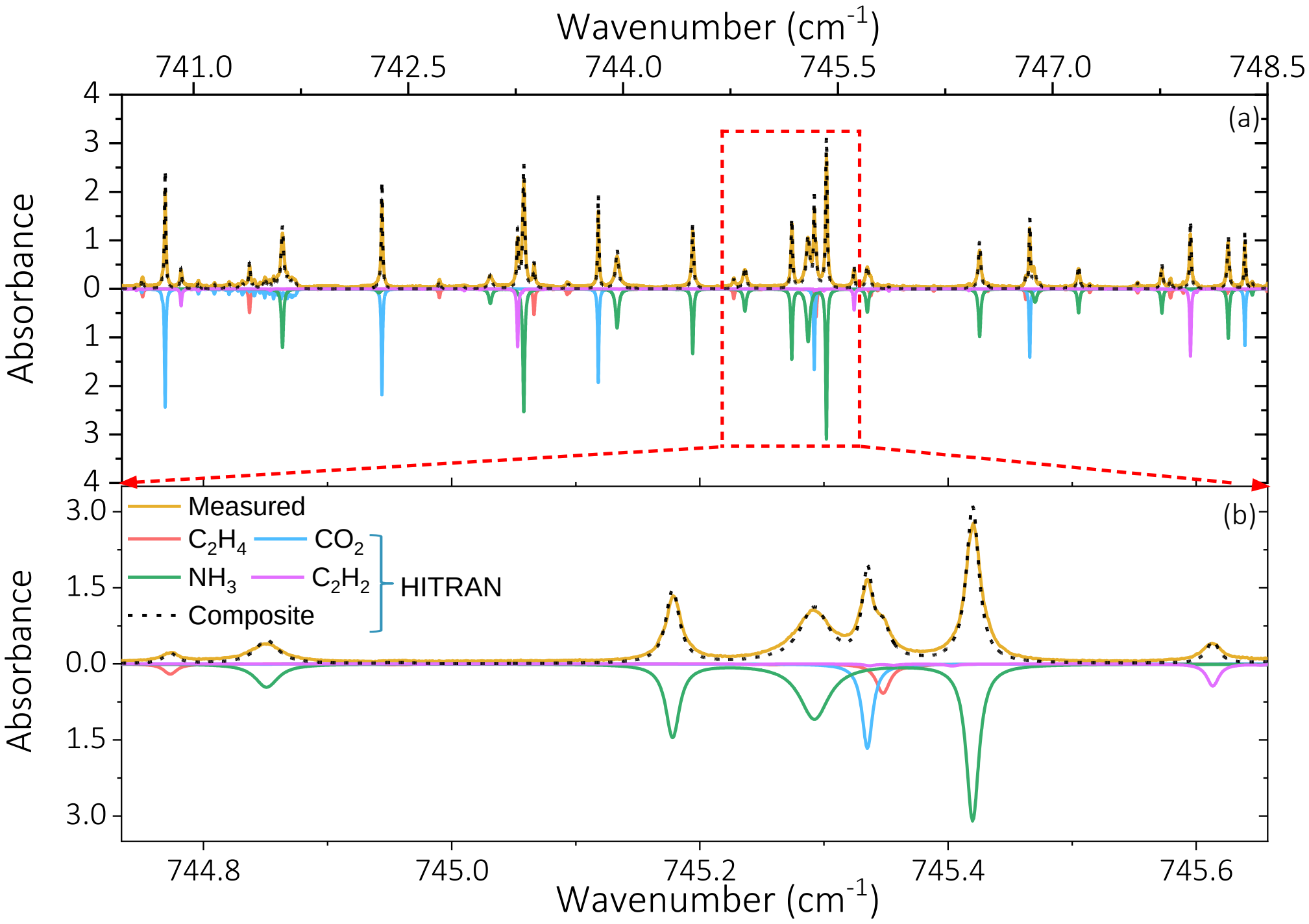

Fig. 5. (a) Top panel: measured (yellow) and simulated (HITRAN2020 [38], black dashed lines) absorption spectrum of the four-species mixture ($C_2H_4$, $NH_3$, $CO_2$ and $C_2H_2$) at 50 Torr and 295 K for a 23-cm pathlength over 740.5 – 748.5 cm$^{-1}$; bottom panel: simulated individual spectra

of the four species at the same conditions using HITRAN2020 [38]. (b) Close-up view of absorption features over 744.73 – 745.66 $cm^{-1}$.

Even in the relatively narrow spectral span of Fig. 5(a), numerous strong spectral lines (> 100) of the four species are observed, some of them being blended together. Figure 5(b) highlights a spectral segment where eight strong spectral lines of the four species emerge over a narrow range of ∼0.93 $cm^{-1}$ scanned in ∼130 ms. It is worth noting that while the integrated absorbance of the simulated composite spectrum matches that of the measured spectrum, the line shapes in the two spectra are somewhat different. This is due to the fact that the bath gas in the simulated HITRAN spectra is assumed to be air while the measured spectrum corresponds to a non-diluted mixture of the four species with unknown collisional broadening coefficients. In particular, $NH_3$ is most likely the highest contributor to the spectral discrepancies, for two reasons: first, the self–broadening coefficient of $NH_3$ ($\gamma_{\mathrm{NH_3-NH_3}}$) in this spectral region is 4 – 6 times higher than that of $\gamma_{\mathrm{X-air}}$ (the pressure broadening coefficient between air and any of the four species in the mixture); second, since the release of HITRAN2016 onwards, $NH_3$ transitions with quantum numbers $J'' > 8$ have been assigned a default value of $\gamma_{\mathrm{self}}$ = 0.5 $cm^{-1}$/atm [39], whereas for transitions with $J'' \leq 8$, an empirical polynomial given by Nemtchinov et al. [40] is used for broadening coefficients. The default value of 0.5 $cm^{-1}$/atm can cause a departure from the prediction of the empirical polynomial by a factor of ∼2, especially for low-$K$ transitions, (e.g., for a transition in the $P$-branch of the $\nu_2$ band with $J''$ = 9 and $K''$ = 1, HITRAN2020 lists $\gamma_{\mathrm{NH_3-NH_3}} =$ 0.5 $cm^{-1}$/atm while the Nemtchinov polynomial [40] gives $\gamma_{\mathrm{NH_3-NH_3}} =$ 0.237 $cm^{-1}$/atm). Among the eighteen $NH_3$ transitions depicted in Fig. 5(a), only two are assigned $J'' \leq 8$, which highlights the potential of the developed DFG laser in updating the pressure broadening coefficients of $NH_3$ for high-$J''$ transitions.

Most $CO_2$ transitions in Fig. 5(a) correspond to $Q$- and $R$-branches of the hot band $\nu_1 + \nu_2 - 2\nu_2$ ($l$ = 2) [41]. The other transitions correspond mostly to the fundamental $\nu_{10}$ ($CH_2$ rocking), $\nu_2$ (symmetric bending), and $\nu_5$ (CH bending) bands of $C_2H_4$, $NH_3$, and $C_2H_2$, respectively. These are the strongest fundamental bands of these three species. Nevertheless, high-resolution measurements of spectral line parameters of these species are scarce in this spectral region, and most of those listed in the HITRAN2020 database come from FTIR measurements at low resolution [38]. For instance, the first measurements of self–shifting and self–broadening collisional coefficients of rovibrational lines in the $\nu_{10}$ band of $C_2H_4$ were carried out recently by Ulenikov et al. using an FTIR spectrometer [42].

To showcase the broadband tunability of the developed DFG laser system, we performed high-resolution absorption measurements of the $\nu_{10}$ band of $C_2H_4$ over the 706.5 – 767.0 $cm^{-1}$ range (13.04 – 14.15 μm) at a pressure of 150 Torr. Figure 6(a) shows the measured $C_2H_4$ spectrum as compared to a simulated spectrum using the HITRAN2020 database [38]. The agreement is remarkable. Due to the constraint of the DFG phase-matching bandwidth, the spectrum was piecewise measured at every ∼6 $cm^{-1}$ interval. Ten intervals were needed to cover the 706.5 – 767.0 $cm^{-1}$ spectral range; each segment is shown with a particular color on the top panel of Fig. 6(a). The $CO_2$ laser was kept fixed at a wavelength of 9.552 μm and at a repetition rate of 10 kHz while the start and end pump wavelengths were chosen to address the entire target idler range. Figure 6(b) shows a close-up view of a single scan segment over 737.2 – 743.3 $cm^{-1}$ for better visibility, while Fig. 6(c) displays the corresponding residuals calculated as 100 × (simulated – measured). The residuals are affected by spikes in correspondence of the absorption peaks that we mainly attribute to the uncertainty in the absorption pathlength and to imprecisions in the wavelength axis calibration.

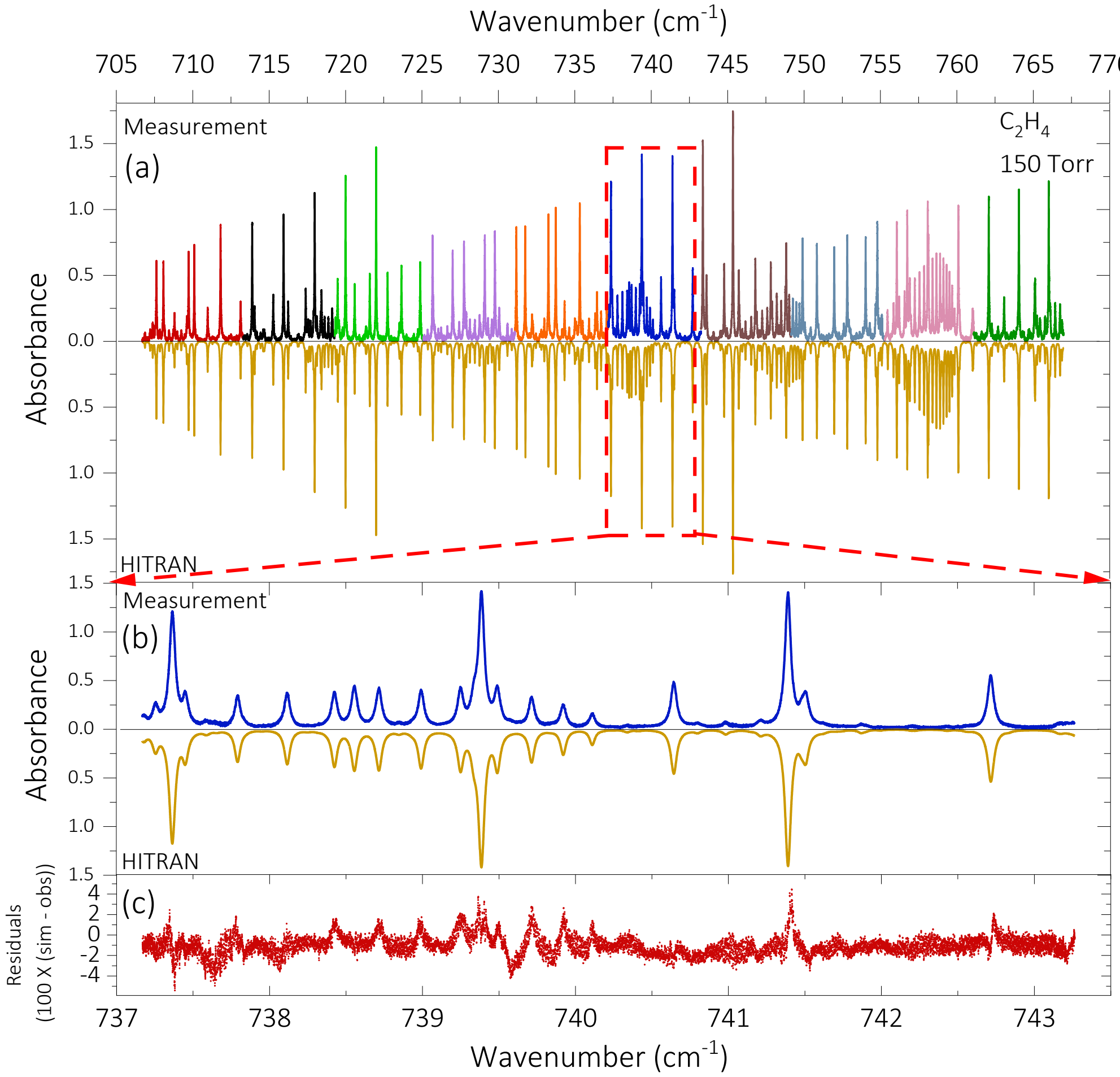

Fig. 6. (a) Measured (top panel) and simulated (HITRAN2020, bottom panel) absorbance spectra of $C_2H_4$ for a 23-cm pathlength over 706.5 – 767.0 $cm^{-1}$ (13.04 – 14.15 μm) at a pressure of 150 Torr and at 295 K. The different colors refer to the different (ten) scan segments adopted to cover the entire range. The phase-matching condition was readjusted from segment to segment. The simulated spectrum of $C_2H_4$ was obtained from the HITRAN2020 database [38]. (b) Close-up view of a single scan segment over 737.2 – 743.3 $cm^{-1}$. (c) Residuals calculated as 100 × (simulated – observed) spectrum for the segment shown in (b).

To evaluate the high-resolution performance of our DFG laser system, we employed it to measure self-broadening coefficients ($\gamma_{self}$) of the six spectral lines of $C_2H_4$ shown in Fig. 4(a). Each of those spectral lines is the sum of a pair of transitions having overlapping linecenters [38]. In fact, each pair (line) includes two transitions which share the same self–broadening coefficient and linecenter, except for the line centered at 740.645 $cm^{-1}$; for that particular line, theoretical calculations (adopted in HITRAN2020 [38]) predict a slight shift of $5.19\times10^{-4}$ $cm^{-1}$ in the linecenter between the two corresponding transitions, as opposed to experimental results which report identical linecenters for both transitions (740.645013 $cm^{-1}$) [42,43]. We will ignore the theoretically predicted shift in our analysis, and we will label the line with the listed HITRAN2020 linecenter (740.645156 $cm^{-1}$) which is closer to the reported empirical results [38]. The linecenters and quantum numbers of the selected lines are listed in Table 1.

We measured the absorption spectrum of the selected lines of $C_2H_4$ over 739.4 – 741.5 $cm^{-1}$ at various pressures between 5 and 50 Torr at 295 K. The $CO_2$ gas laser wavelength was set to 9.552 μm with a pulse repetition rate of 10 kHz while the EC-QCL wavelength was scanned over 5.5916 – 5.5982 μm. Absorption line shapes were fitted using the Voigt line-shape function. Figures 7($a_1$) and ($b_1$) show the measured absorbance spectra along with the line-shape fitting of the spectral line centered at 741.39 $cm^{-1}$ at a pressure of 5 and 50 Torr, respectively. The residuals are shown in Figs. 7 ($a_2$) and ($b_2$) and are calculated as 100 × (fit – measured). The gull-wing shaped residuals near the peak (within 6% of the peak absorbance which is typical for Voigt-profile fitting [44]) is attributed to the fact that the Voigt profile does not account for the effects of collisional narrowing and speed-dependent broadening which are relatively significant at lower pressures (i.e., lower number densities) [44]. This also explains the higher residuals at 5 Torr as compared to 50 Torr, as shown in Fig. 7($a_2$) and ($b_2$).

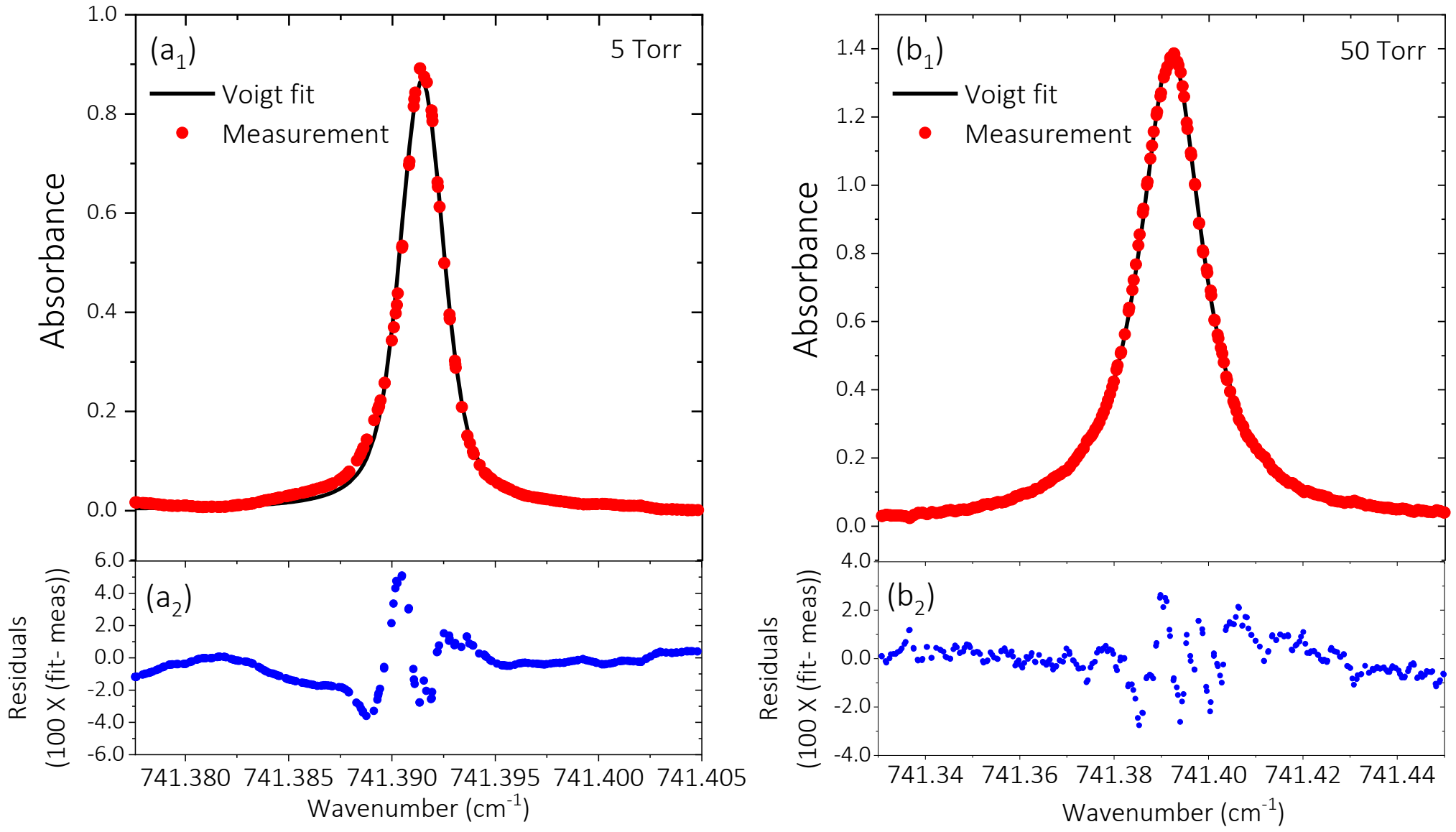

Fig. 7. Measured absorbance spectra (red circles) and fitted Voigt profiles (black lines) for the line centered at 741.39 $cm^{-1}$, at 295 K for a 23-cm pathlength: ($a_1$) at a pressure of 5 Torr and ($b_1$) at 50 Torr. ($a_2$) and ($b_2$): Residuals calculated as [100 × (fitted value – measured value)].

The collisional halfwidths at half maximum (collisional HWHM) obtained from the Voigt fitting of the six lines are plotted as a function of gas pressure in Figs. 8(a-f). The red lines represent the linear best fit between the collisional halfwidths and the gas pressure, according to the relation: HWHM [$cm^{-1}$] = $\gamma_{self}$ [$cm^{-1}atm^{-1}$] × $P$[atm]. Therefore, the slopes of the red lines in Figs. 8(a-f) represent the self-broadening coefficients of the six spectral lines. The plotted confidence bands represent the slope uncertainties based both on the standard deviation of residuals and on the shown vertical and horizontal error bars, corresponding to the uncertainties in the fitted collisional halfwidths and in the pressure (±0.12% of pressure reading), respectively. The vertical intercepts of the six fitted regression lines fall between $-2.2\times10^{-4}$ and $1.8\times10^{-4}$ $cm^{-1}$. Table 1 lists the obtained self–broadening coefficients and compares them against the listed values in HITRAN2020. The reported uncertainty in HITRAN2020 of these line parameters is 5 − 10% (HITRAN uncertainty code is 5) [38]. As pointed out in the last column of Table 1, all measured broadening coefficients agree with those in HITRAN2020 within the claimed uncertainty range in HITRAN2020, except for the fourth spectral line centered at 740.11 $cm^{-1}$ (−10.48% difference with HITRAN2020) that anyway respects a 1-sigma agreement if one considers the combined uncertainty of the two values.

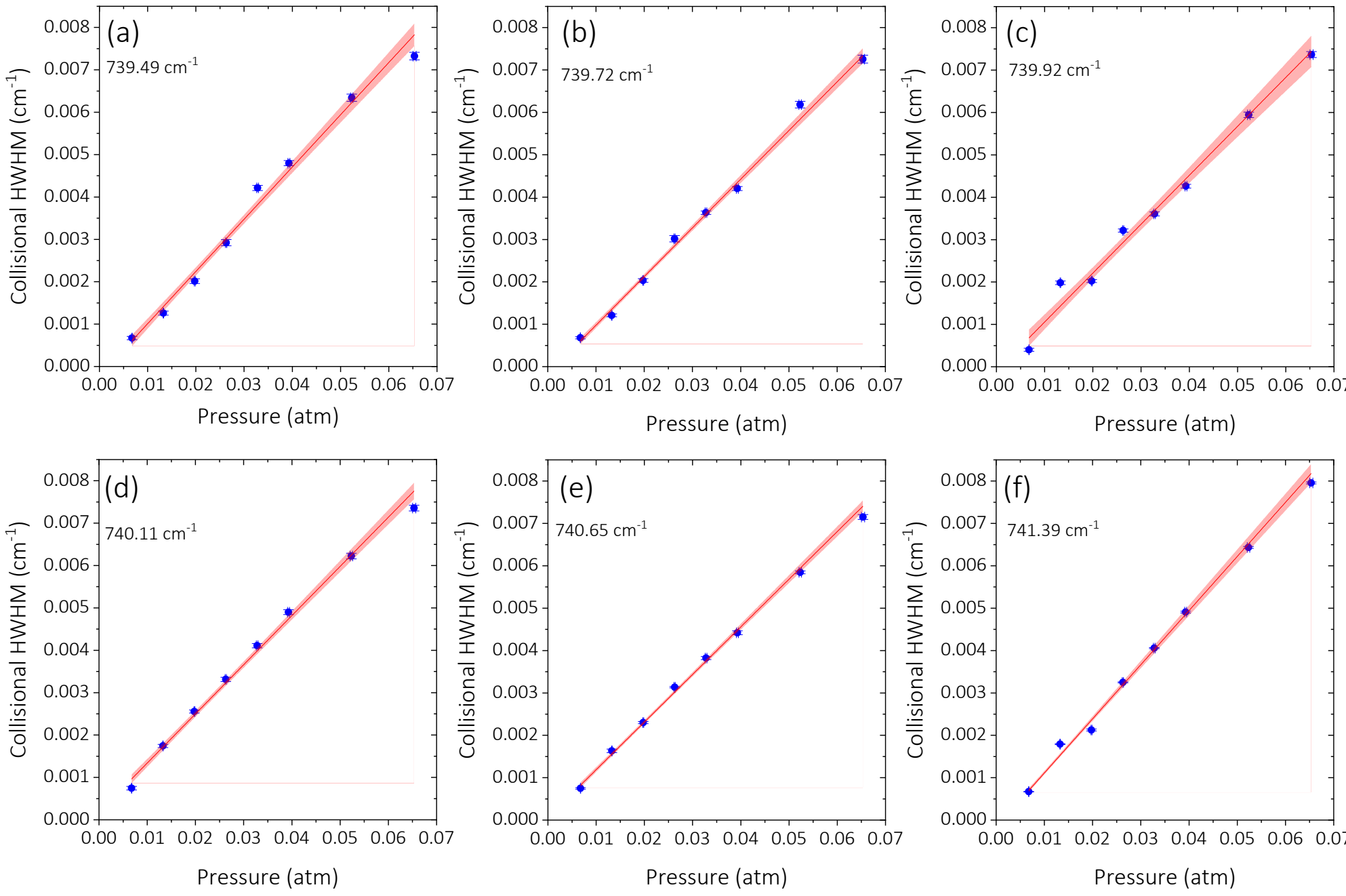

Fig. 8. Collisional half-widths (blue symbols) of pure $C_2H_4$ over $5 - 50$ Torr at 295 K; red lines in panels (a-f) are the linear best fits for the selected lines shown in Fig. 4(a), with red confidence bands representing slope uncertainties; the shown label in each panel denotes the linecenter.

**Table 1. Self–broadening coefficients of selected lines of $C_2H_4$ at 295 K.**

| $\nu_0$ [cm$^{-1}$][a] | Transition quantum numbers[b] | $\gamma_{\text{self}}$ [cm$^{-1}$atm$^{-1}$] (This work)[c] | $\gamma_{\text{self}}$ [cm$^{-1}$atm$^{-1}$] (HITRAN2020)[d] | $\delta$%[e] |
|---|---|---|---|---|
| 739.491333 | (13 9 5) – (13 10 4) | 0.1233(53) | 0.1150(58) | -7.17 |
| | (13 9 4) – (13 10 3) | | | |
| 739.715212 | (12 9 3) – (12 10 2) | 0.1149(33) | 0.1130(57) | -1.64 |
| | (12 9 4) – (12 10 3) | | | |
| 739.921945 | (11 9 3) – (11 10 2) | 0.1154(77) | 0.1110(56) | -3.99 |
| | (11 9 2) – (11 10 1) | | | |
| 740.111492 | (10 9 2) – (10 10 1) | 0.1160(40) | 0.1050(53) | -10.48 |
| | (10 9 1) – (10 10 0) | | | |
| 740.645156 | (14 6 8) – (15 7 9) | 0.1124(26) | 0.1190(60) | 5.53 |
| | (14 6 9) – (15 7 8) | | | |
| 741.391503 | (9 7 2) – (10 8 3) | 0.1277(36) | 0.1180(59) | -8.24 |
| | (9 7 3) – (10 8 2) | | | |

[a]Linecenter from HITRAN2020 database [38].

[b]Each spectral line is the sum of a pair of overlapping transitions whose quantum numbers $\left((J', K_a', K_c') - (J'', K_a'', K_c'')\right)$ are shown in this column.

[c]Values in parentheses are 1σ experimental standard errors.

[d]$\gamma_{\text{self}}$ from HITRAN2020 database [38]. Values in parentheses correspond to reported minimum uncertainty of 5%.

[e]Percentage difference between measured and HITRAN2020: $100 \times (\gamma_{\text{self,HITRAN}} - \gamma_{\text{self,measured}})/\gamma_{\text{self,HITRAN}}$.

## 4. Summary

We have proposed an approach for high-resolution spectroscopy in the long-wavelength part of the molecular fingerprint region that relies on widely tunable DFG laser pulses at kHz repetition rates and peak power at the tens of μW level. The DFG nonlinear laser source is in-house built and covers an unprecedented ultra-wide range from 11.58 μm (863.56 $cm^{-1}$) to 15 μm (666.67 $cm^{-1}$) that intercepts intense absorption features of a wealth of molecular species, from both vibrational and bending modes. The wide spectral coverage comes from the broad tunability of both pump and signal lasers: the former can be finely mode-hop-free tuned around 5.6 μm, while the signal laser can lase over multiple lines in a kHz-pulsed mode over the 9.23 – 10.476 μm range, resulting in a semi-continuous idler laser emission. The idler radiation exhibits a linewidth of ~2.3 MHz and an average output power exceeding 30 μW, which is sufficient to bring into saturation liquid-nitrogen-cooled detectors and achieve high signal-to-noise ratios. We tested the ability of the developed system for multispecies detection by measuring the absorption spectra of a mixture of $C_2H_4$, $CO_2$, $NH_3$ and $C_2H_2$ over 740.5 – 748.5 $cm^{-1}$ and by retrieving their respective mole fractions. Furthermore, we showcased the broadband tunability of our spectrometer by measuring the absorbance spectrum of $C_2H_4$ over 13.04 – 14.15 μm and by leveraging the high resolution of the spectrometer to obtain self–broadening coefficients of six $C_2H_4$ spectral lines in agreement with HITRAN2020. We believe that our approach for high-resolution broadband mid-IR spectroscopy has great potential for analytical applications and fundamental molecular studies. Furthermore, the DFG laser system can be automated in such a way that the crystal position is continuously adjusted during the pump wavelength scans to preserve phase matching while varying the pump wavelength.

## Funding

This work was funded by the Office of Sponsored Research (OSR) at King Abdullah University of Science and Technology (KAUST).

## Disclosures

The authors declare no conflicts of interest.